%% file: 00paper.tex
\documentclass[runningheads]{llncs}
\usepackage[T1]{fontenc}
\usepackage{graphicx}

\usepackage{url}

\usepackage{booktabs} 
\usepackage{color}
\usepackage{makecell}
\usepackage[square,comma,numbers,sort&compress,sectionbib]{natbib}
\usepackage{amssymb}
\usepackage{enumitem}

\usepackage{fancyhdr}

\fancypagestyle{specialfooter}{%
  \fancyhf{}
  
  \fancyfoot[C]{\vspace{1cm} \hspace*{-.2\textwidth}\parbox{1.4\textwidth}{\small This is the author's version of the work. It is posted here for your personal use.}}
}

\begin{document}
\title{SimLab: A Platform for Simulation-based Evaluation of Conversational Information Access Systems}
\titlerunning{SimLab: A Platform for Simulation-based Evaluation of CIA Systems}
\author{Nolwenn Bernard\thanks{Work done while at the University of Stavanger, Norway.}\inst{1} %
\and
Sharath Chandra Etagi Suresh\inst{2} %
\and
Krisztian Balog\inst{3} %
\and ChengXiang Zhai\inst{2} %
}
\authorrunning{Bernard et al.}
\institute{TH Köln, Köln, Germany\\ \email{nolwenn.bernard@th-koeln.de} \and
University of Illinois at Urbana-Champaign, Urbana, IL, USA \and
University of Stavanger, Stavanger, Norway}
\maketitle              %
\thispagestyle{specialfooter}

\begin{abstract}
Progress in conversational information access (CIA) systems has been hindered by the difficulty of evaluating such systems with reproducible experiments. While user simulation offers a promising solution, the lack of infrastructure and tooling to support this evaluation paradigm remains a significant barrier. To address this gap, we introduce SimLab, the first cloud-based platform providing a centralized solution for the community to benchmark both conversational systems and user simulators in a controlled and reproducible setting.
We articulate the requirements for such a platform and propose a general infrastructure to meet them. We then present the design and implementation of an initial version of SimLab and showcase its features through an initial simulation-based evaluation task in conversational movie recommendation.  
Furthermore, we discuss the platform's sustainability and future opportunities for development, inviting the community to drive further progress in the fields of CIA and user simulation.

\end{abstract}

\input{ecir2026-simlab-01}
\input{ecir2026-simlab-02}
\input{ecir2026-simlab-03}

\input{ecir2026-simlab-04}

\input{ecir2026-simlab-05}
\input{ecir2026-simlab-06}

\subsubsection{\ackname} This research was supported by the Norwegian Research Center for AI Innovation, NorwAI (Research Council of Norway, proj. nr. 309834).

\bibliographystyle{splncs04nat}
\bibliography{ecir2026-simlab}

\end{document}

%% file: ecir2026-simlab-01.tex
\section{Introduction}

\begin{figure}[t]
    \centering \includegraphics[keepaspectratio,scale=0.8]{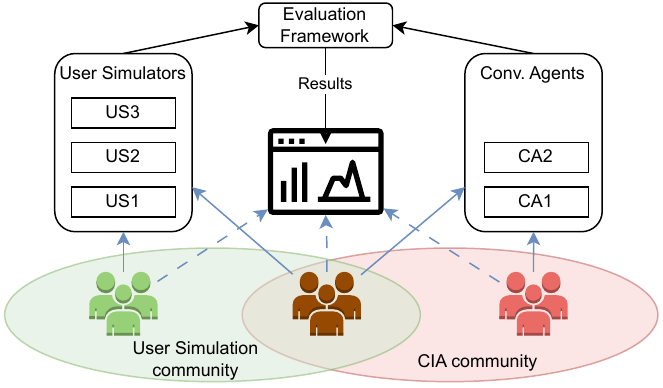}
    \caption{Vision for SimLab including the different stakeholders and resources involved in the simulation-based evaluation of conversational agents. The straight blue arrows represent the submissions of conversational agents and user simulators, while the dashed blue arrows represent access to the results.}
    \label{fig:vision}
\end{figure}

Evaluation is critical to both research and development, as it allows us to assess a system's performance over time and to compare it with other systems. Flawed evaluation inevitably hinders research progress and leads to the deployment of suboptimal systems. Recently, conversational agents have emerged as an alternative to traditional search engines and recommender systems for accessing information~\citep{Culpepper:2018:SIGIRForum,Zamani:2023:FnTIR}.
They allow users to disclose and iteratively refine information needs and provide direct feedback via a sequence of turns---features not inherently supported by traditional systems.
However, their evaluation remains an open challenge in part due to their highly interactive nature and inadequacy of traditional evaluation methods and metrics~\citep{Balog:2024:FnTIR,Zamani:2023:FnTIR}. 

Simulation-based evaluation has been identified as a promising approach to address limitations related to interactivity and reproducibility in evaluation~\citep{Balog:2024:FnTIR}. Indeed, this evaluation paradigm involves substituting real users with simulated ones to interact with the system in a controlled and reproducible environment. However, its adoption is still limited, partly due to the scarcity of adapted resources~\citep{Balog:2024:FnTIR,Breuer:2025:SIGIRForum}.
In fact, using user simulators implies the reliance on a software library or toolkit that can generate interactions with the agent under evaluation.
While some initiatives are emerging to address these limitations, e.g.,~\citep{Azzopardi:2024:SIGIRAP,Afzali:2023:WSDM}, they often have a narrow application range. That is, they are commonly tailored to specific information access tasks, such as interacting with a ranked list of search results or getting item recommendations from a conversational agent, where only a limited set of user and agent actions are defined and supported. This hinders their generalizability to other interactive information access tasks. 

We introduce SimLab to foster wider adoption of simulation-based evaluation among researchers and practitioners, and to ultimately facilitate advancements in conversational assistant research and development. Its main objective is to provide a centralized platform to benchmark both conversational information access (CIA) systems and user simulators for various application domains and tasks through simulation-based evaluation. Figure~\ref{fig:vision} illustrates the general vision of SimLab as a potentially self-sustaining ecosystem.
Here, different stakeholders (e.g., researchers, students, and benchmark organizers) from various communities can collaborate, each contributing by sharing their resources, while simultaneously benefiting from the resources shared by others. For example, a researcher interested in user simulation can focus on developing a new user simulator and evaluate it on a specific task, without needing to also develop conversational agents, and vice versa. 

To achieve this, we begin by identifying the requirements for the platform to ensure it addresses the needs of the community. These requirements are informed by the strengths and limitations of existing solutions for benchmarking (interactive) information access systems, e.g.,~\citep{Frobe:2023:SIGIR,Azzopardi:2024:SIGIRAP}. For example, SimLab should support online evaluation while being flexible enough to accommodate a wide range of systems (conversational agents and user simulators) for various tasks and domains. To meet these requirements, we present a cloud-based platform and its key components, and demonstrate its functionality via an initial evaluation task focused on conversational recommender systems in the movie domain. Finally, we discuss SimLab's opportunities and challenges for the community.

In summary, this work presents SimLab, a cloud-based platform that facilitates simulation-based evaluation of CIA agents as well as the development of user simulators, detailing the requirements that guided its design and implementation. The first release of SimLab, including an initial evaluation task, is available at \mbox{\url{https://iai-group.github.io/simlab/}}, with the corresponding implementation at \mbox{\url{https://github.com/iai-group/simlab}}.

%% file: ecir2026-simlab-02.tex
\section{Background}
\label{sec:related}

The evaluation of conversational information access agents remains a challenging and open task~\citep{Zamani:2023:FnTIR,Balog:2024:FnTIR}. 
In this section, we first review traditional evaluation methods for CIA agents. We then discuss how simulation-based evaluation as a promising approach to address their limitations. Finally, we explore the evaluation-as-a-service paradigm and its potential to support this approach.

\subsection{Traditional Evaluation Methods}

Traditionally, CIA agents are evaluated using three main methodologies: (1) online evaluation, (2) user studies, and (3) offline evaluation~\citep{Zamani:2023:FnTIR,Gao:2023:Book}.
Online evaluation deploys a system to real users and collects feedback from them. While providing the most reliable results, it is expensive, cumbersome, and requires a large, active user pool. User studies recruit participants to evaluate the system at a smaller scale, hence reducing the costs. However, the involvement of humans, whose behavior cannot be fully controlled, makes experiments nearly impossible to reproduce.
Offline evaluation removes the human factor by using automatic metrics over static test collections. This approach is the foundation for many benchmarks that have driven progress in information access, from international campaigns like TREC and CLEF to collections like BEIR~\citep{Thakur:2021:NeurIPS}. This approach has also been extended to CIA-specific benchmarks, such as the TREC Conversational Assistance Track (CAsT)~\citep{Dalton:2020:arXiv} and the Interactive Knowledge Assistance Track (iKAT)~\citep{Aliannejadi:2024:arXiv}.

Despite their value, these traditional methods---particularly offline and user-study-based benchmarks---have significant limitations. Standard offline metrics (e.g., Recall@N, BLEU) often disregard the dialogue structure and correlate poorly with user satisfaction~\citep{Liu:2016:EMNLP,Jiang:2016:CHIIR,Bernard:2025:arXiv}. Furthermore, many benchmarks abstract away the interactive nature of CIA by using pre-defined user trajectories and static run submissions rather than dynamic, end-to-end systems (with some exceptions, like iKAT 2025). This, combined with a lack of publicly available system implementations, creates serious reproducibility challenges. To address some of these metric-related flaws, the rise of large language models (LLMs) has recently led to the development of LLM-based evaluators to measure various aspects of CIA agents~\citep{Hashemi:2024:ACL,Huang:2024:arXiv,Aliannejadi:2024:arXiv}.

\subsection{Simulation-based Evaluation}

Simulation-based evaluation has been identified as a promising approach to evaluate interactive systems, including CIA agents, and is (re)gaining interest~\citep{Azzopardi:2011:SIGIRForum,Balog:2024:FnTIR,Breuer:2025:SIGIRForum}. A user simulator may be viewed as an intelligent agent designed to mimic the behavior of real users interacting with the system under evaluation. Therefore, the simulated interactions can be leveraged to assess the system's performance. 
One of the main advantages of this approach is the ability to control the behavior of simulated users, which is not possible with human subjects, thus facilitating the reproducibility of the experiments.
While simulation-based evaluation alleviates key limitations of traditional evaluation methods, such as the need for human participants and the lack of interactivity, it also introduces its own challenges~\citep{Zamani:2023:FnTIR,Breuer:2025:SIGIRForum,Balog:2024:FnTIR}, including the realism of simulators and standard definitions of success criteria.
It is also important to realize that standard evaluation measures are essentially simulators, albeit often simplistic ones (e.g., NDCG@k assumes a user that examines the first $k$ results in sequential order and derives higher utility from relevant results returned at higher rank positions).  The goal, therefore, is to support evaluation with more realistic simulators. 

In practice, simulation-based evaluation of CIA agents is still in its early stages. It may be explained by the complexity of the task, the skepticism over results obtained using simulated users, which is due in part to the lack of validation methods/metrics for user simulators~\citep{Breuer:2025:SIGIRForum}, and the scarcity of open resources~\citep{Balog:2024:FnTIR}. This work focuses on the latter by proposing an open cloud-based platform.
Various toolkits and frameworks have been developed to facilitate simulation-based evaluation of CIA agents~\citep{Azzopardi:2024:SIGIRAP,Kiesel:2024:CLEF,Afzali:2023:WSDM,Fu:2024:UMCIR,Wang:2023:EMNLP,Huang:2024:arXiv}. 
While these resources are valuable starting points, they often have a narrow application range; for example, SimIIR 3.0~\citep{Azzopardi:2024:SIGIRAP} is specifically designed for search engines and CSSs, and UserSimCRS~\citep{Afzali:2023:WSDM} is tailored for conversational recommender systems. Furthermore, they often leave the burden of reproducing and adapting (existing) conversational systems to researchers.

\subsection{Evaluation-as-a-Service}

Unlike traditional benchmarks, the Evaluation-as-a-Service (EaaS) paradigm~\citep{Hopfgartner:2018:JDIQ,Lin:2013:SIGIRForum} considers algorithms as pieces of software that can be evaluated against unseen data rather than submitting system outputs generated offline (``runs''). This presents several advantages, such as the centralization of the resources and automation of evaluation, thus, facilitating its standardization and reproducibility. However, EaaS also presents challenges, primarily maintenance, operationalization, and cost.
An example implementation of EaaS is the TIRA Integrated Research Architecture (TIRA)~\citep{Frobe:2023:ECIR}. Its main objective is to support the organization of shared task events, where the systems submitted by the participants can be stored, maintained, and re-used for future evaluations, facilitating the reproducibility of experiments. The IR Experiment Platform~\citep{Frobe:2023:SIGIR} is an example of successful integration of TIRA for evaluating information retrieval systems.

This paradigm appears well suited for SimLab, which aims to support simula\-tion-based evaluation of CIA agents at scale. In fact, CIA agents and user simulators can be viewed as software components that can be executed together in a controlled environment to perform the evaluation. Therefore, it embraces the interactive nature of the systems under evaluation.

%% file: ecir2026-simlab-03.tex
\section{SimLab}
\label{sec:simlab}

\begin{figure*}[t]
    \centering
    \includegraphics[keepaspectratio,width=\columnwidth]{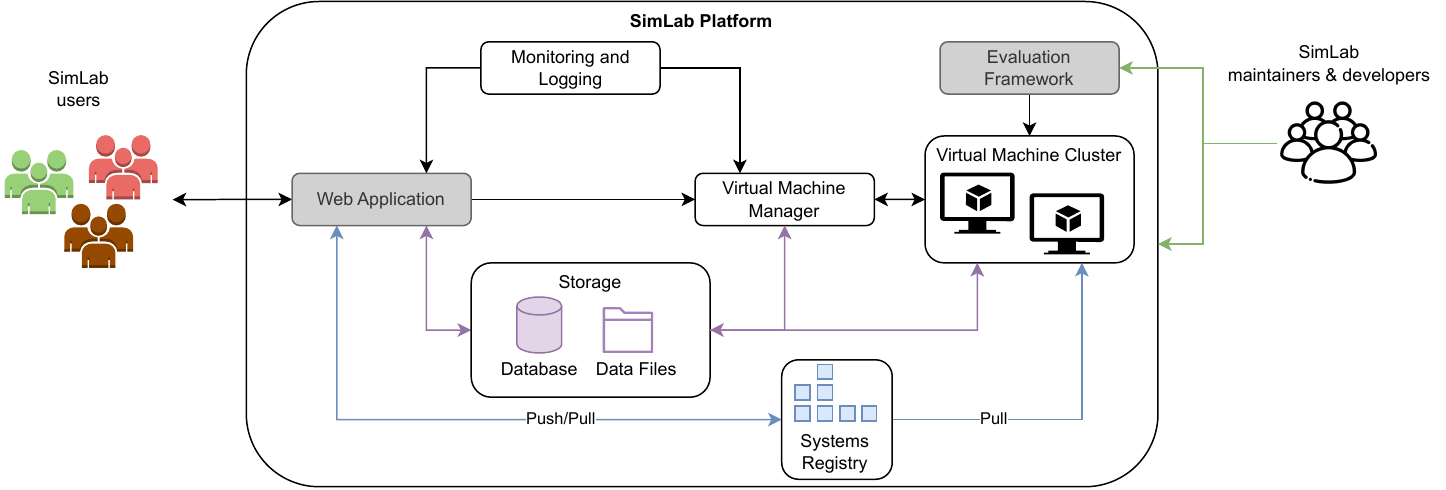}
    \caption{Overview of the SimLab platform. Purple and blue arrows represent data and system flow respectively, while the green arrows indicate contributions to the platform by maintainers and developers. Grey boxes denote the components implemented by us.}
    \label{fig:environment}
\end{figure*}

This work introduces SimLab, a benchmark platform facilitating simulation-based evaluation of conversational information access systems for various tasks and application domains.
It can provide rich insights for different objectives, such as analyzing the strengths and weaknesses of systems,  evaluating novel methods and algorithms, and comparing system variants for industrial applications.
Hence, it can support various stakeholders, including system developers, researchers, and educators, who are interested in user simulation and/or CIA. 

In this section, we first discuss the requirements for SimLab %
and then describe the proposed cloud-based platform, which is depicted in Fig.~\ref{fig:environment}. 

\subsection{Platform Requirements}
\label{sec:simlab:req}

The simulation-based evaluation process involves at least two dialogue participants: a CIA agent and a user simulator. Therefore, the platform must support online evaluation by collecting conversations between these participants, enabling the automatic evaluation of new systems (i.e., CIA agent or simulator). Consequently, the platform must be flexible enough to run a wide range of CIA agents and simulators while orchestrating their interactions.
As the platform is intended to serve a community of users, it should be scalable to dynamically accommodate evolving traffic needs. Additionally, it must be portable to simplify migration to different deployment environments.
Finally, the platform should be intuitive to use and extensible, making it easy for the community to adopt and contribute to its development.

In summary, we identify the following main requirements:
\begin{itemize}
    \item[R1] \textbf{Online Evaluation Support}:  Provides a solution for running CIA agents and user simulators, and collecting their conversations.
    \item[R2] \textbf{Flexibility}:  Supports a wide range of CIA agents and user simulators, including different frameworks and programming languages.
    \item[R3] \textbf{Scalability and Portability}: Handles varying traffic loads and is deployable in diverse environments with minimal effort.
    \item[R4] \textbf{Intuitive Use}: Offers an easy-to-use interface for conducting evaluation experiments.
    \item[R5] \textbf{Extensibility}:  Allows for the addition of new components and functionalities to enhance the platform over time.
\end{itemize}

\subsection{The SimLab Platform}

Next, we present the SimLab platform and discuss how its components address each requirement.

\subsubsection{Online Evaluation Support (R1)}

To support online evaluation, we introduce the platform's core component: the \emph{Evaluation Framework}. It defines the concepts and processes for simulation-based evaluation, including the definition of evaluation tasks, metrics, and the communication protocol between CIA agents and user simulators. 
The communication protocol is the key element supporting the collection of conversations, which serve as the basis for evaluation. It defines the structure of the messages and their metadata exchanged between the two systems. SimLab imposes no restrictions on system implementation; the only requirement is adherence to this communication protocol.

\begin{figure}[t]
    \centering    \includegraphics[keepaspectratio,width=\columnwidth]{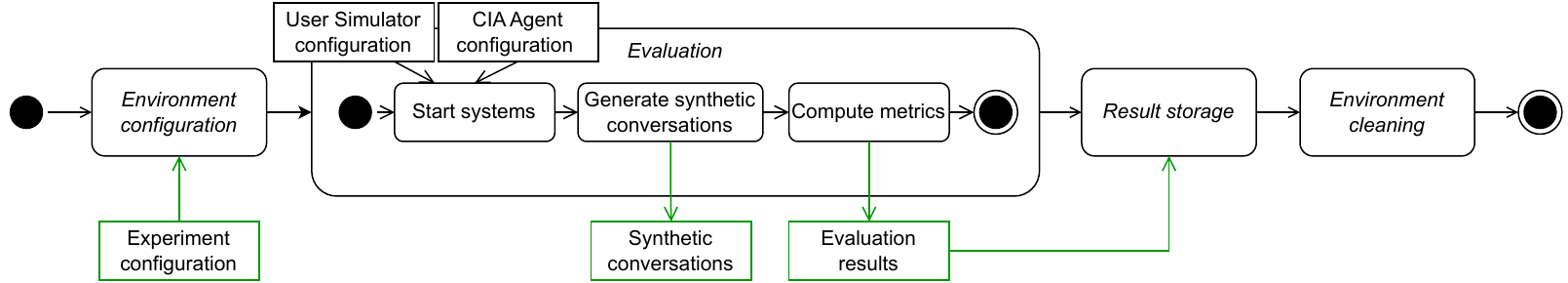}
    \caption{UML activity diagram of the simulation-based evaluation process. The experiment data, highlighted in green boxes, is saved in the \emph{Storage}. It is assumed that the systems are available in the \emph{Systems Registry}.}
    \label{fig:eval-uml}
\end{figure}

The simulation-based evaluation process, depicted in Fig.~\ref{fig:eval-uml}, consists of four main stages:
\begin{enumerate}[topsep=2pt]
    \item \emph{Environment Configuration}: The experiment environment is prepared by instantiating the evaluation task, along with its associated metrics and evaluation information needs. The configuration is stored in the \emph{Storage} for reproducibility purposes.
    \item \emph{Evaluation}: For each experiment, a CIA agent and simulator pair are started within the experiment environment. Conversations between them are collected (based on evaluation information needs) and stored in the \emph{Storage}. Metrics are then computed from these conversations.
    \item \emph{Result Storage}: The evaluation results for each system pair are saved in the \emph{Storage}.
    \item \emph{Environment Cleaning}: The experiment environment is deleted to prepare for the next experiment.
\end{enumerate}

\subsubsection{Flexibility (R2)}

To support a wide range of systems---both CIA agents and simulators---they are treated as containerized software. This allows systems to be stored and maintained in the \emph{Systems Registry}, making them easily reusable for future evaluations.
Furthermore, containers are self-contained and lightweight, allowing developers to implement their systems in their preferred environment with minimal constraints for integration with SimLab. This streamlines the development process of systems. 

\subsubsection{Scalability and Portability (R3)}

SimLab is designed as a cloud-based platform to benefit from the scalability and elasticity of cloud computing~\citep{Mell:2011:NIST}. In this context, a dynamic \emph{Virtual Machine Cluster}, a set of virtual machines (VMs), to execute the simulation-based evaluation process. Depending on the cluster size, multiple experiments can run in parallel, increasing the platform's robustness to traffic variations.
To manage the VM workload, we introduce a \emph{Virtual Machine Manager} that maintains a queue of evaluation experiments submitted to SimLab. It strategically allocates experiments to available VMs, queueing them if none are free. It also dynamically scales the number of VMs based on the current workload to ensure efficient resource utilization. 
Additionally, SimLab integrates \emph{Monitoring and Logging} systems to provide insights into efficiency, traffic, and potential errors via customizable dashboards. These insights can inform configuration or hardware decisions, such as increasing the VM cluster size to reduce waiting times if the experiment queue is consistently growing. The platform's components are standard in cloud environments, easing portability to different cloud providers with minimal effort. 

\subsubsection{Intuitive Use (R4)}

To provide a user-friendly experiment interface, we developed a \emph{Web Application} that serves as the main user interface. It combines a graphical user interface and an API, connecting users to the evaluation process and public parts of the \emph{Storage} (e.g., results, experiment configurations, and training data). The web application allows users to log in, submit systems (CIA agents and simulators), manage their experiments, and access the storage. The user interface is designed to be highly intuitive, lowering the barrier to entry for the community.

\subsubsection{Extensibility (R5)}

Both the platform and web application are designed to be generic, following principles of modular and object-oriented programming.
It enables the platform to be easily extended with new metrics, tasks, and application domains, as long as they conform to the high-level definitions of these concepts. The web application can also be extended with new visual components and routes to support new features.
In practice, SimLab is released as an open-source project, allowing the community to directly contribute new features, bug fixes, and improvements via the code repository.

%% file: ecir2026-simlab-04.tex
\section{Implementation and Deployment}
\label{sec:imp}

SimLab's implementation and deployment utilize a containerized, cloud-native architecture to ensure scalability, monitoring, and automated orchestration. This section details the implementation of SimLab's key components. For each component, we provide a brief overview of the chosen tools, frameworks, and technologies, justifying our selections. Finally, we discuss our cloud provider selection and present estimated deployment costs for the platform. 

\begin{table*}[t]
    \small
    \centering
    \caption{Communication protocol endpoints per system type to perform simulation-based evaluation in SimLab.}
    \label{tab:endpoints}
    \begin{tabular}{l@{~~}|p{4.8cm}|c|c}
        \hline
        \textbf{Endpoint (\texttt{POST})} & \textbf{Description} & \textbf{CIA Agent} & \textbf{User Sim.} \\ \hline
        \texttt{/configure} & Configure the system with parameters & \checkmark & \checkmark \\ \hline
        \texttt{/receive\_utterance} & Receive an utterance from the other system and generate a response & \checkmark & \checkmark \\ \hline
        \texttt{/set\_information\_need} & Set the information need for the conversation & $\times$ & \checkmark \\ \hline
        \texttt{/get\_information\_need} & Get the current information need & $\times$ & \checkmark \\ \hline
    \end{tabular}
\end{table*}

\paragraph{Evaluation Framework}
The evaluation framework is developed in Python. It leverages the DialogueKit~\citep{Afzali:2023:WSDM} library, which provides fundamental concepts for CIA, such as conversational participants and dialogue orchestration. The general concepts related to simulation-based evaluation are implemented as extensible modules. 
For the communication protocol between CIA agents and simulators, we rely on HTTP requests based on a minimal set of predefined endpoints (listed in Table~\ref{tab:endpoints}) and data formats. Using HTTP requests for the interactions allows for easy integration with various programming languages and frameworks.

\paragraph{Web Application}
The backend is developed using Flask,\footnote{\url{https://flask.palletsprojects.com/en/stable/}} chosen for its lightweight nature and ability to handle RESTful APIs efficiently. Flask provides a modular and minimalist approach, making it suitable for microservice-based deployments. The frontend is implemented using React, which offers a component-based architecture, enabling reusable UI elements and efficient state management.

\paragraph{Database}
SimLab relies on MongoDB\footnote{\url{https://www.mongodb.com}} as the primary database, chosen over traditional relational databases for its flexibility and scalability. 
Unlike relational databases, which typically rely on vertical scaling, MongoDB supports horizontal scaling through sharding, enabling efficient distribution of large datasets across multiple nodes.
Additionally, MongoDB's schemaless architecture allows for dynamic data structures (stored as JSON-like documents) that evolve with application needs, making it well-suited for research environments and AI-driven applications. It simplifies data retrieval and manipulation, aligning well with the requirements of modern cloud-based applications.

\paragraph{Systems Registry}
Systems are stored in a dedicated Docker\footnote{\url{https://www.docker.com}} registry, enabling version control and tagging. Using Docker images to package systems ensures that they are isolated and can easily be deployed in the experiment environment. This choice is motivated by the wide adoption of Docker as a containerization platform and its compatibility with various cloud providers. 

\paragraph{Virtual Machine Manager}
For automated provisioning and scaling of virtual machines, Jenkins\footnote{\url{https://www.jenkins.io}} plays a critical role. It manages the lifecycle of worker nodes based on workload demand, ensuring that computational resources are dynamically allocated. Jenkins also provisions jobs that require containerized execution, facilitating concurrent job processing while optimizing cloud resource usage.

\paragraph{Monitoring and Logging}
We utilize Grafana\footnote{\url{https://grafana.com}} and Prometheus\footnote{\url{https://prometheus.io}} to monitor metrics such as CPU usage, memory consumption, and job execution times. Prometheus scrapes time-series data from our cloud infrastructure, while Grafana provides real-time visualization and alerting mechanisms. These tools are chosen for their wide adoption in the industry and active community support.

\begin{table}[t]
    \small
    \centering
    \caption{Estimation of the monthly costs for the different components of the platform, including their specifications, on different cloud providers.}
    \label{tab:costs}
    \begin{tabular}{l|c|c|c}
        \hline
        \textbf{Component} & \textbf{AWS} & \textbf{Azure} & \textbf{GCP} \\ \hline
        \emph{Web Application} + \emph{Monitoring \& Logging} & & & \\
        \hspace{1cm} 1 VM (2 vCPU, 4 GB RAM) & \$37 & \$38 & \$35 \\ \hline
        \emph{Virtual Machine Manager} & & & \\
        \hspace{1cm} 1 VM (2 vCPU, 4 GB RAM) & \$37 & \$38 & \$35 \\ \hline
        \emph{Virtual Machine Cluster} & & & \\
        \hspace{1cm} 1 VM (4 vCPU, 16 GB RAM) & \$103 & \$104 & \$101 \\ \hline
        \emph{Data Files Storage} (1 TB) & \$23  & \$20 & \$22 \\ \hline
        \emph{Database} (20 GB MongoDB) & \$120 & \$125 & \$110 \\ \hline
        \emph{Systems Registry} (Docker Registry) & \$5 & \$5 & \$5 \\ \hline
        \hline
        SimLab Platform & \$325 & \$330 & \$308 \\
        \hline
    \end{tabular}
\end{table}

\paragraph{Cloud Provider}
For the first version of SimLab, we selected Google Cloud Platform (GCP) as our cloud provider. This decision was based on a preliminary cost analysis comparing pricing from three main providers for the different resources required to operationalize SimLab; see Table~\ref{tab:costs}. Note that the costs can increase depending on the size of the VM cluster and the configuration chosen for the different resources.
We acknowledge that a multi-cloud strategy could further reduce the costs, yet, it also increases the complexity in network configuration and service interoperability. Therefore, we leave this for future work.

%% file: ecir2026-simlab-05.tex
\section{First Use Case: Conversational Movie Recommendation}

In this section, we showcase SimLab's features through an initial evaluation task: evaluating CIA agents in the movie domain.
We use movie attributes, such as genre, keywords, and actors, from the MovieLens 32M dataset~\citep{Harper:2015:TIIT} to create information needs. An information need comprises a set of constraints defining the desired movie and a set of attributes the user wants to know about. For example, a simulator might look for a 2009 romantic comedy movie while being curious about its duration and associated keywords~\citep{Zhang:2020:KDD}. The evaluation assesses utility using success rate (proportion of successful conversations based on the assessment of a zero-shot classifier) and dialogue capabilities using understanding and consistency aspects from the fine-grained evaluation of dialog metric (FED)~\citep{Mehri:2020:SIGDIAL}.

\begin{figure*}[t]
    \centering
    \includegraphics[keepaspectratio,width=\columnwidth]{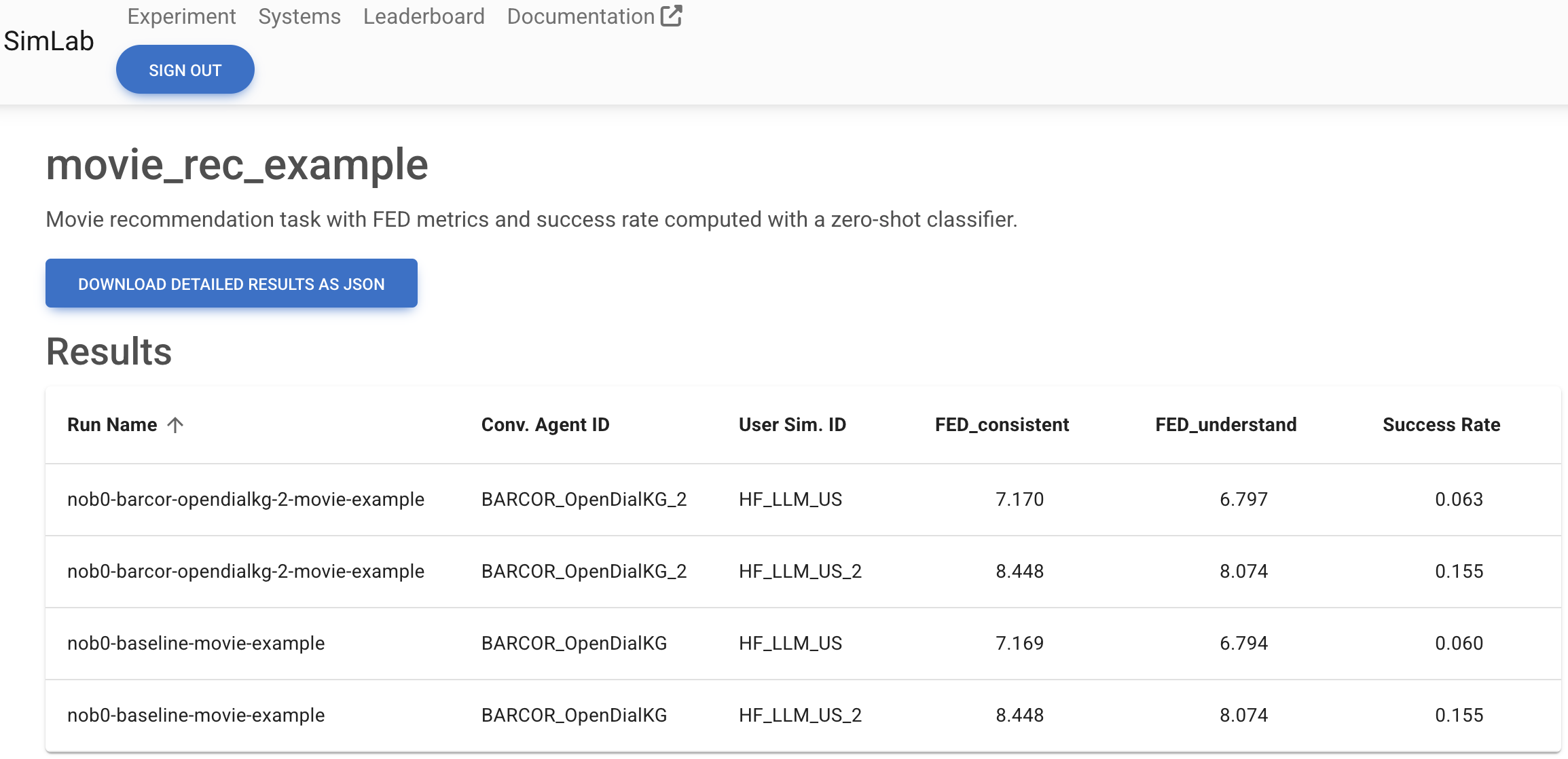}
    \caption{Leaderboard page for the movie recommendation task. The results for two runs are placed in a table below the ``Download results'' button.}
    \label{fig:leaderboard}
\end{figure*}

To showcase SimLab's basic features, we upload Docker images of conversational agents (adapted from the iEvaLM repository~\citep{Wang:2023:EMNLP}) and simulators. More specifically, we perform experiments using the BARCOR~\citep{Wang:2022:arXiv} OpenDialKG CRS and a naive simulator based on the \texttt{SmolLM2-135M-Instruct} model~\citep{Allal:2025:arXiv} from the Transformers library~\cite{Wolf:2020:EMNLP}, with each having two versions (1.0 and 2.0). Synthetic conversations are generated based on a set of information needs. %
These conversations are then evaluated and the results are displayed on the leaderboard page, as depicted in Fig.~\ref{fig:leaderboard}. The page presents the results in a tabular format, where each row represents a CRS agent and simulator combination. The table is sortable by the different metrics to facilitate systems comparison. The leaderboard page also provides an option to download detailed evaluation results as a JSON file, including metric statistics, for further analysis.

This experiment enables the investigation of various research questions such as: How does the simulator's response generation prompt affect the CRS agent's ability to provide successful recommendations? Do the modifications applied to BARCOR OpenDialKG 2.0 improve its dialogue capabilities compared to the previous version? 
However, the focus of this paper remains on the platform itself, not on answering these specific research questions.

%% file: ecir2026-simlab-06.tex
\section{Discussion and Conclusion}
\label{sec:discussion}

In this work, we have introduced SimLab, a cloud-based platform facilitating simulation-based evaluation of conversational information access systems. It aims to provide a centralized solution to benchmark both conversational systems and user simulators in a controlled and reproducible environment.
SimLab presents opportunities for various stakeholders, including researchers, practitioners, and educators in the fields of user simulation and CIA. We conclude by discussing some of these opportunities and their associated challenges.

\paragraph{Accelerating Research.}
SimLab facilitates experimentation in the fields of user simulation and CIA, allowing stakeholders to investigate new metrics tailored to CIA scenarios as well as different types of systems. For example, one can compare the performance of conversational agents with different interaction styles (e.g., template- vs. generative-based) and their generalizability to different tasks and application domains. It promotes a continuous improvement and refinement cycle in which progress is driven by findings from community experiments.
The main challenge is supporting a wide range of scenarios and systems, while minimizing the constraints on the users and remaining user-friendly.
Moreover, the data collected during the experiments can be used to create new synthetic datasets for future research, although their release should be done with caution to avoid disseminating poor quality, harmful and/or biased data~\citep{Soudani:2024:arXiv}.

\paragraph{Facilitating Reproducibility and Data Analysis.}
SimLab archives all the experiments and metadata to facilitate reproducibility and in-depth data analysis. It eliminates the need to re-implement, fix, or adapt baseline systems as the systems and experiment configurations are readily available. Furthermore, analysis of saved results can yield insights into the strengths and weaknesses of the different systems. The first version of SimLab has a basic leaderboard that ranks the pairs of conversational agents and simulators based on their performances on a specific task. However, there are opportunities to extend this leaderboard with additional features that can automate in-depth analysis of the results. For example, an aggregation algorithm could automatically compare the performance of different systems across multiple tasks and domains.

\paragraph{Supporting Education.}
SimLab has the potential to support education, particularly project-based learning, in CIA and user simulation.
It enables learners to work on specific information access tasks by experimenting with different systems and comparing their results.
For example, SimLab can be used to set up hands-on assignments where students work on a specific task, compare results, and share insights. Challenges to effectively support learning include integrating a discussion forum where learners can interact with each other and an intelligent tutor to provide personalized feedback to the learners.

\paragraph{Democratizing Simulation-based Evaluation.}
We acknowledge that user simulators can be imperfect, explaining the skepticism regarding simulation-based evaluation~\citep{Breuer:2025:SIGIRForum}; yet, we argue that all existing evaluation methods assume some kind of simulation. 
Research has shown that results from such imperfect evaluation can still be useful for identifying strengths and weaknesses of conversational agents in a controlled environment (see, e.g.,~\citep{Voorhees:2022:arXiv}).  Moreover, these findings can serve as a filter for underperforming agents before moving to human evaluation, allowing efficient use of scarce and expensive human resources. 
To contribute to the acceptance of simulation-based evaluation, we envision the integration of human benchmarks and feedback with conversational agents in future versions of SimLab. This would facilitate automatic validation of simulators by assessing their abilities to approximate human performance~\citep{Bernard:2024:ICTIR}.

\paragraph{Building a Self-Sustaining Ecosystem.} 
As mentioned in our vision, the long-term success of SimLab depends on fostering a self-sustaining community. The platform is designed to create a ``virtuous cycle:'' researchers contribute CIA agents to benchmark them, which in turn enriches the pool of agents for simulator developers to test against. Similarly, new simulators provide new evaluation perspectives for agent developers. A key challenge will be incentivizing these initial contributions. We plan to mitigate this by collaborating with shared tasks (like TREC) and integrating SimLab into university curricula, seeding the platform with an initial critical mass of systems, tasks, and users.